# Electrical transport across metal/two-dimensional carbon junctions: Edge versus side contacts

Yihong Wu[1], Ying Wang[1], Jiayi Wang[2], Miao Zhou[3], Aihua Zhang[3], Chun Zhang[3,4], Yanjing Yang[1,5], Younan Hua[5] & Baoxi Xu[6]

[1]Department of Electrical and Computer Engineering, National University of Singapore, 4 Engineering Drive 3, Singapore 117576, Singapore

[2]NUS Graduate School for Integrative Sciences and Engineering, Centre for Life Sciences, #05-01, 28 Medical Drive, Singapore 117456, Singapore

[3]Department of Physics, National University of Singapore, 2 Science Drive 3, Singapore 117542, Singapore

[4]Department of Chemistry, National University of Singapore, 3 Science Drive 3, Singapore 117543, Singapore

[5]GLOBALFOUNDRIES, Singapore Pte Ltd, 60 Woodlands Industrial Park D Street 2 Singapore 738406, Singapore

[6]Data Storage Institute, 5 Engineering Drive 1, National University of Singapore, Singapore 117608, Singapore

Metal/two-dimensional carbon junctions are characterized by using a nanoprobe in an ultrahigh vacuum environment. Significant differences were found in bias voltage (V) dependence of differential conductance (dI/dV) between edge- and side-contact; the former exhibits a clear linear relationship (i.e., $dI/dV \propto V$), whereas the latter is characterized by a nonlinear dependence, $dI/dV \propto V^{3/2}$. Theoretical calculations confirm the experimental results, which are due to the robust two-dimensional nature of the carbon materials under study. Our work demonstrates the importance of contact geometry in graphene-based electronic devices.

## I. INTRODUCTION

An ideal graphene is a monatomic layer of carbon atoms arranged on a honeycomb lattice. This unique lattice structure leads to the formation of quasi-relativistic low-energy excitations near the K points at the corners of the first Brillouin zone; the quasi-particles are chiral and massless Dirac fermions with the electrons and holes degenerated at the Dirac points (or K points).[1-5] These unique electronic structures in turn give rise to a number of peculiar physical properties of graphene distinguishing it from conventional two-dimensional electron gas systems, some of which are desirable for high-frequency electronics applications.[6-8] Comparing to graphene itself, however, our understanding of metal (M) - graphene (G) contact is still far from complete, which may eventually limit the performance of graphene-based electronic devices.[9] Experimental and theoretical studies have shown that both the nature of atomic bonding at the metal-graphene interface and band structure of graphene near the Fermi level play crucial roles in determining the transport properties of M-G junctions.[9-13] Here, we demonstrate that, in addition to the type of contact materials and corresponding nature of atomic bonding between metal and graphene, the contact geometry, i.e., an edge- or side-contact,[14] also plays an important role in electron transport across the M-G junctions.

## II. EXPERIMENTS

Compared to the side-contact, which is routinely employed for electrical transport measurement of graphene or graphene electronic devices, it remains a great challenge to form a pure edge-contact with graphene without touching its surface due to its ultra small thickness. In order to overcome this difficulty, in this work, we use a position-controllable nanoprobe to form edge-contacts with two different types of two-dimensional (2D) carbons, both of which have free-standing edges above the substrate surface on which they are either grown or placed. The first type of 2D carbon is so-called carbon nanowalls (CNWs),[15,16] which are curved carbon nano-sheets grown almost vertically on flat substrates. Figure 1(a) shows the schematic of an edge-contact between 2D carbon and a tungsten (W) probe. In order to form a reliable contact, our measurements were performed using an Omicron ultrahigh vacuum (UHV) system with a base pressure in the range of $3-8\times10^{-11}$ Torr. Equipped in the UHV system are a scanning electron microscope (SEM) and four independently controllable nanoprobes with auto-approaching capability, which allows for position-specific measurements with good reproducibility. Figure 1(b) shows an example of an edge-contact formed by a tungsten probe and a piece of CNW. The size of the contact is determined mainly by the thickness of the CNW which is about one to several nanometers at the edge,[15,16] and is adjustable through monitoring the zero-bias contact resistance. The second type of 2D carbon is obtained *in-situ* through mechanical exfoliation of highly ordered pyrolytic graphite (HOPG) by using a large-size probe which itself also forms a low-resistance contact with HOPG during the subsequent electrical measurements (see Figure 1(c)). As can be seen in this figure, an edge-contact can be readily formed between a second W probe and the edge of an exfoliated 2D carbon sheets. The precise positioning of probe allows for formation of contacts between the probe and different points of the edge. Again, the actual contact size can be adjusted manually through monitoring the contact resistance. Alternatively, an edge-contact can also be formed by probing directly the edge of a small piece of HOPG flake placed with an off-angle from the flat surface of a substrate holder.

All the electrical measurements were performed using a standard lock-in technique at room temperature. During the measurements, one of the probes was used to form a low-resistance contact and the other is adjusted manually to have a different contact resistance. The automatic approaching function helps to make reliable and reproducible contact without damaging the sample and the probe, unless the substrate is an insulator in which case the first probe has to be sacrificed by being pressed manually on the sample.

## III. RESULTS AND DISCUSSION

### A. Structural properties of CNWs

Figure 2 (a) shows the typical SEM image of CNWs which consist of networks of submicron-sized graphite sheets having a typical height of ~ 2 μm and a width less than 1 μm.[15,16] Also seen in the figure is the W probe with a diameter of about 150 nm. Figure 2 (b) shows the SEM image of the same sample after it was coated with a thin layer of Fe with a nominal thickness of 30 nm. As it will be discussed shortly, this sample serves as a control for the bare CNW samples. Figure 2 (c) shows the high-resolution transmission electron microscopy (HRTEM) image of a single piece of CNW. The image demonstrates clearly the layered graphitic structure. The thickness of the graphite sheets varies from sample to sample and sheet to sheet; it is typically in the range of one to several nanometers.[17] Some of the CNWs are just a few layers (e.g., the inset shows the image of a CNW with a thickness of 2-3 layers), which makes it straightforward to form a point-contact at the edge even with a relatively large probe. The average interlayer spacing of CNWs is about 0.336 nm, which is slightly larger than that of graphite.[17] This is presumably caused by the local curvature in the CNWs.

The CNWs can be grown on any type of substrates provided that substrates can sustain a temperature of 650-700°C.[15-17] In this work, the CNWs have been grown on both $SiO_2$/Si and Cu substrates. The CNWs grown on the Cu substrates can be peeled off easily due to weak bonding between carbon and Cu. After peeling-off, a thin layer of carbon is often found present on the Cu substrate near the unpeeled region (see Figure 3(a)). In order to understand quantitatively the chemical composition and structure of this thin layer of material, Auger element mapping has been carried out for three different regions of a same sample: region with CNWs (i), exposed Cu substrate covered by a thin layer of carbon (ii), and completely exposed Cu substrate (iii). The Auger spectra taken from regions (i) and (ii) resembles closely the Auger spectra of few layer graphene reported by Xu *at al*.,[18] while that of region (iii) is dominated by peaks associated with copper. Based on Auger spectra, element mappings are obtained for carbon and copper and the results are shown in Figures 3(b) and (c), respectively. The Auger results are in good agreement with the growth mechanism of CNW or carbon nanosheets reported in literature, i.e., a thin layer of flat graphite sheets is formed on the substrate at the initial stage, followed by the growth of vertical nanosheets.[19,20] This provides a convenient way to have both flat and vertical carbon nanosheets on a same substrate through peeling off and flipping over a small portion of CNW, as shown in the SEM image of Figure 3(d). Auger element mapping confirms that both region A (back surface of flipped over region) and region B (original CNW) of the sample shown in Figure 3(d) are few layer graphene sheets. This unique type of sample allows for direct comparison between side- and edge-contact by simply positioning the 2$^{nd}$ probe on different regions of the sample, with the other probe, which is firmed pressed onto the substrate to form a low-resistance contact, remained at the same position.

**B. dI/dV curves for edge-contacts with bare CNWs**

Figure 4 (a) shows the dI/dV curves as a function of bias voltage obtained by first forming an ohmic contact with CNWs grown on $SiO_2$/Si using one of the probes and then varying the sample-probe distance of the second probe. In order to avoid causing damage to the nanowalls, the second probe was first placed on the nanowalls through the auto-approaching function and then the sample-probe distance was adjusted manually through monitoring the zero-bias resistance (ZBR) using a small AC current. The dI/dV versus bias voltage curves were then recorded by sweeping the DC bias current within a predefined range and measuring the dI/dV using the lock-in technique. The different curves shown in Figure 4(a) are corresponding to contacts with different ZBR values ranging from 4.6 to 26.1 k$\Omega$. For the sake of clarity, all the curves other than the one with highest ZBR (the bottom curve) are shifted upward in the figure. The symbols are the measurement results and solid-lines are linear fittings to the experimental data. When both probes were in ohmic contact with the nanowalls, the resistance measured was in the range of 200-600 $\Omega$, for a probe distance of 1 – 20 μm. In this case, the dI/dV curve is almost independent of the bias voltage. Therefore, the measured conductance shown in Figure 4(a) is dominantly due to the second CNW-probe contact. As can be seen from the HRTEM image shown in Figure 2(c), although the CNWs are few layer graphene sheets, the electrical contact was presumably formed with the layer (or layers) of highest height. Under the ballistic transport approximation, the conductance of one transverse mode of a graphene point-contact is $4e^2/h$, corresponding to a resistance of 6.45 k$\Omega$. Therefore, during the measurements, the ZBR has been varied in the range of ~3 - 30 kΩ. Although there is no clear definition between the point-contact and tunneling regimes, the point-contact regime can be considered as being in the range where ZBR is close to the resistance quanta of one conduction channel. When the ZBR reaches a value which is several times that of the resistance quanta, it is more appropriate to treat the contact as being in the tunneling regime. As can be seen from Figure 4(a), the differential conductance is linear to the bias voltage for all the ZBR values that have been measured. The experiments have been repeated on different CNW samples and also at many different locations for a same sample, which exhibited excellent reproducibility.

**C. dI/dV curves for edge-contacts with Fe-coated CNWs**

After a series of measurements were completed on bare CNWs, the sample was *in-situ* coated with a thin layer of Fe in the preparation chamber and then transferred back to the measurement chamber for performing the same series of electrical measurements without breaking the vacuum. As it is shown in Figure 2(b), the coating is uniform and has a nominal thickness of about 30 nm. The dI/dV curves for the Fe-coated sample are shown in Figure 4(b). In a sharp contrast to the case of bare CNWs, now the dI/dV curves exhibit a well-defined parabolic shape. Compared with the bare CNW sample, it is generally more difficult to form a stable contact between the probe and Fe-coated CNWs. Therefore, after several runs of measurements, the probe has to be lifted for re-establishing a new contact. This resulted in dI/dV curves with different rate of change with respect to the bias voltage. Nevertheless, all the experimental data (symbols) are fitted well using the relation $dI/dV \propto V^2$, with V being the probe-sample bias voltage. This kind of parabolic dI/dV curve is normally obtained in metal-insulator-metal tunnel junction.[21] The good agreement between theoretical and experimental curves for the Fe-coated sample confirms unambiguously that the linear curves shown in Fig. 4(a) are due to the CNW-W edge-contact.

**D. Comparison of edge- and side-contacts formed with CNWs**

The same measurements were then performed on CNWs grown on a Cu substrate, which allows for partial peeling-off of CNW from the substrate. Note that the conducting substrate does not affect the measurement results because the resistance measured mainly comes from the contact with a larger resistance. Figures 5(a) and (b) show the dI/dV curves obtained from the back surface (after peeling-off from the substrate) and front end of CNWs, corresponding to region A and B of Figure 3(d), respectively. The experimental data (symbols) are fitted well to the relation $dI/dV \propto V^{3/2}$ for (a) and $dI/dV \propto V$ for (b), respectively. Again, for the sake of clarity, all the curves other than the one with highest ZBR (the bottom curve) are shifted upward in the figure. As the measurements on regions A and B of Figure 3(d) correspond to side- and edge-contact, respectively, the results demonstrate that the dI/dV curves are indeed dependent on the relative orientation between the probe (or more accurately, current direction) and base plane of the carbon lattice.

**E. Comparison of side- and edge-contact formed with HOPG and exfoliated graphene sheets**

In order to further confirm the results shown in Figures 4 and 5, we repeated the same series of experiments on HOPG and exfoliated graphene sheets. In order to reduce the influence of surface contaminants, the top layer of HOPG was *in-situ* peeled off using a probe. Compared to CNWs, it was found that it is generally more difficult to form reliable side-contact with thick HOPG plates due to its flatness and hardness. However, once a stable contact is formed, the measurement results are reproducible at different locations on the HOPG surface. On the other hand, it is relatively easy to form a stable contact from the edges for both thick flakes and few-layer graphene sheets [see Figure 1(c)]. In the former case, although the flake is thick, contact is presumably only formed at point (or points) with highest protrusions. Therefore, the actually contact only occurs at the edge of graphene sheets. On the other hand, in the latter case, it is determined dominantly by the thickness of graphene sheets. The dI/dV curves obtained from both the side- and edge-contacts are shown in Figures 6(a) and 6(b), respectively. Again, the dI/dV curves in Figure 6(a) are fitted well by the relation $dI/dV \propto V^{3/2}$ for all ZBR values. On the other hand, the dI/dV curves in Figure 6(b) follow closely a linear-dependence on the bias voltage, as is the case in Figures 4(a) and 5(b). In addition to measurements carried out for contacts with ZBRs in the range of 4.7-13.1 k$\Omega$, we have also measured the dI/dV of edge-contacts with a ZBR in the range of 256 - 476 $\Omega$. As shown in Figure 7, the dI/dV still exhibits a linear-dependence with the bias voltage. This shows clearly that the bias-dependence of dI/dV is mainly determined by the intrinsic property of graphene rather than the W probe.

In order to compare the results with Fe-coated CNWs, we have also performed similar experiments on Fe-coated HOPG. As with the case of Fe-coated CNWs, the Fe coating has a nominal thickness of 30 nm. The experiments have been repeated on two samples with separately coated Fe layer and on different experiment runs. In both cases, the dI/dV-V curves exhibit a well-defined parabolic shape. Figure 8 shows the results obtained from one of the samples at different zero-bias resistance values. The results are reproducible at different locations of the same sample.

## F. Theoretical calculation of dI/dV curves

The experimental results may be understood intuitively by considering the difference in relative orientation of the Fermi surface of graphene (with a disk shape) with respect to the current direction for two different contacts, as illustrated schematically in Figure 9 (side-contact, a and edge-contact, b). The difference is obvious because in the case of side-contact, graphene does not have a wavevector component in the current direction, which is not the case for edge-contact. For a more quantitative understanding, we have carried out first-principles calculations for the edge-contact and for the side-contact we have used an analytical model to estimate the dI/dV curves. To simplify the problem, we ignore the scattering of electrons by the interface, and assume that every channel inside graphene can accept electrons from the metal probe with a unity transmission. We first consider the edge-contact. The electron transport of this case can be modeled by graphene sandwiched between two metal electrodes as shown in Figure 10(a). In the figure, shadowed areas L and R denote two contact regions between graphene and metal electrodes. If neglecting the scattering of electrons at metal-graphene interfaces, the electron transport of this case can be solved by first-principles methods combining density functional theory and Green's functions' techniques[22] with two semi-infinite contact regions, L and R. Calculations were performed using the computational software ATK.[23] In Figure 10(b), we show the calculated dI/dV as a function of bias voltage. Clearly, in this case, dI/dV is a linear function of $V$, which is actually expected since in this case, the transport is governed by density of states (DOS) of graphene that is linear at the vicinity of Fermi energy. Contact resistance will introduce the scattering of electrons at interfaces, and consequently decrease the current. If the scattering is energy independent, dI/dV will still approximately be linear as a function of V but with a decreased slope.

We next consider the case that the probe axis is perpendicular to the base plane of HOPG, i.e., surface-contact. The electron transport process in this case is illustrated in Figure 11, which is different from that of edge-contact in a sense that in this case, in order to tunnel into graphene, electrons in the probe have to possess a finite momentum in the direction normal to the graphene plane, $k_\perp$. The external bias applied is approximately shared equally by the shift of Fermi levels of both the tip and graphene, then one has $\frac{eV}{2} = \frac{\hbar^2 (k_t - k_{tF})^2}{2m^*} \approx \frac{\hbar^2 k_{tF} k_t'}{m^*} = \frac{\hbar^2 k_{tF}}{m^*} \sqrt{k_{t\perp}'^2 + k_{t//}'^2}$, here m* is the effective mass of electrons in the tip, $k_{tF}$ is the Fermi wave vector of the tip, and $k_t' = k_t - k_{tF}$, is the wavevector measured from the Fermi surface, $k_{t//}'$ and $k_{t\perp}'$ are the lateral and longitudinal components of wave vector in the tip, respectively. As shown in the right panel of Figure 10, in this case, we also assume that every state in graphene can accept tunneling of electrons from the tip with a unity transmission. Since the DOS of graphene is given by $\rho(E_G') = \frac{2}{\pi} \frac{|E_G'|}{\hbar^2 v_F^2} = \frac{2}{\pi} \frac{k_{G//}'}{\hbar v_F}$, where $E_G'$ is the energy measured from the Dirac point and $k_{G//}'$ is the lateral wavevector of graphene, number of electrons in the probe that can enter graphene at a particular vertical momentum, $k_{t\perp}'$, can be estimated by $\rho(k_{t\perp}') = \frac{m^* eV}{\pi \hbar^3 v_F k_{tF}} \sqrt{1 - \left(\frac{2\hbar^2 k_{tF} k_{t\perp}'}{m^* eV}\right)^2}$. In deriving this, the metal probe is regarded as a reservoir that has infinite number of transport channels. Then, the average differential conductance can be approximately computed as $\frac{dI}{dV} \propto \frac{1}{k'_{t\perp,\max}} \int_0^{k'_{t\perp,\max}} \rho(k'_{t\perp}) dk'_{t\perp}$, with $k'_{t\perp,\max} = \frac{\sqrt{m^* eV}}{\hbar}$. The integration yields $\frac{dI}{dV} \propto V^{\frac{3}{2}}$.

## IV. CONCLUSIONS

In summary, we have experimentally shown that the electron transport across metal/two-dimensional carbon interface is anisotropic, depending on whether the contact is made from the edge or surface. Experimental results may be understood by taking into account the electron spectrum of both metal and graphene, and also the momentum-resolved DOS of the system. The results are useful for further optimization of metal-graphene contact, a crucial issue pertaining to device applications of graphene.

Acknowledgments

The authors are grateful to P. Xu, S.Y.H. Lua, H.M. Wang, S. S. Kushvaha, C Zhang, and B Wu for their assistance with some of the measurement work. The authors wish to acknowledge funding support from the National Research Foundation of Singapore (YW)(Grant No. NRF-G-CRP 2007- 05 and R-143-000-360-281) and NUS Academic Research Fund (CZ) (Grant No. R-144-000-237-133 and R-144-000-255-112).

Figure captions

FIG.1. Setup of differential conductance measurement: a) illustration of edge-contact between of W probe and 2D carbon; b) SEM image showing experimental realization of edge-contact between W probe and edge of CNW; c) SEM image showing experimental realization of edge-contact between W probe and graphene sheets obtained by *in-situ* exfoliation of HOPG inside the UHV chamber.

FIG.2. a) SEM images of CNWs grown on $SiO_2$/Si substrate (also seen in the image is the W tip used for conductance measurement of edge-contact); b) SEM image of CNW coated with a thin layer of iron with a nominal thickness of 30 nm; c) HRTEM image of a piece of CNW, which demonstrates the high-degree of graphitization of the sample. The inset shows the image of a CNW with a thickness of 2-3 layers.

FIG.3. a) SEM image of CNWs grown on Cu substrate. A portion of CNW has been peeled off by a tweezers. The darker region (ii) near the unpeeled portion is covered by a thin layer of carbon; b) Auger mapping of carbon using peaks associated with graphene sheets of the sample whose SEM image is shown in (a). Graphene sheets were found to exist in both regions (i) and (ii); c*)* copper mapping of the same sample; d) SEM image of CNW sample (B) with a piece of flipped over CNW (A). Auger mapping confirms that region A consists of flat graphene sheets. Portions A and B were used to form side- and edge-contacts, respectively.

FIG.4. a) dI/dV curves as a function of bias voltage for CNW edge-contacts, at different ZBR values (4.6 - 26.1kΩ); b) dI/dV curves as a function of bias voltage for Fe-coated CNW contacted at the edge by W probe, at different values of ZBR (3.7 - 21.9kΩ). The Fe-coating was performed *in-situ* in the same UHV system. For clarity, all the curves other than the one with highest ZBR are shifted upward (same for Figures 5-7). The ZBR values (resistance at the lowest point of the dI/dV curve) for different contacts are listed at the top of the figure in unit of kΩ.

FIG.5. a) dI/dV curves as a function of bias voltage for the back surface of CNWs [region A of Figure 3(d)], at different ZBR values (9.41 - 20.1kΩ), which corresponds to a side-contact. The CNWs were peeled off and flipped over *in-situ* in the same UHV system; b) dI/dV curves as a function of bias voltage for CNW edge-contacts [region B of Figure 3(d)], at different ZBR values (8.35 - 13.1kΩ).

FIG.6. a) dI/dV curves as a function of bias voltage for HOPG surface-contact, at different ZBR values (5.15 - 21.7kΩ). The top layer of HOPG was peeled off *in situ* using one of the probes to avoid the influence of surface contamination; b) dI/dV curves as a function of bias voltage for HOPG edge-contact, at different values of ZBR (4.71 - 13.1 kΩ).

FIG.7. Differential conductance curves of edge-contacts formed with exfoliated graphene sheets in the low contact resistance regime (ZBR ranges from 256 to 476 Ω).

FIG.8. dI/dV curves as a function of bias voltage for Fe-coated HOPG surface, at different ZBR values (0.82 – 1.67kΩ).

FIG.9. Schematic of relative orientation of graphene Fermi surface with respect to the current direction: a) surface-contact, and b) edge-contact. For simplicity, the Fermi surface of metal probe is assumed to have a spherical shape, whereas that of graphene under finite bias is denoted by a circular disk.

FIG.10. a) Theoretical model of graphene sandwiched between two metal electrodes (shadowed areas L and R denote two contact regions); b) calculated dI/dV curve as a function of bias voltage for the in-plane transport with two semi-infinite contact regions. Calculations were done using first principles methods combining DFT and Green's functions' techniques.

FIG.11. Model of vertical electron transport (side-contact): left panel, the sketch of the system for which the axis of the STM tip is normal to graphene plane; right panel, the sketch of the transport process. Electrons with a vertical momentum can enter graphene, and every state in graphene can accept electrons from the probe without scattering.

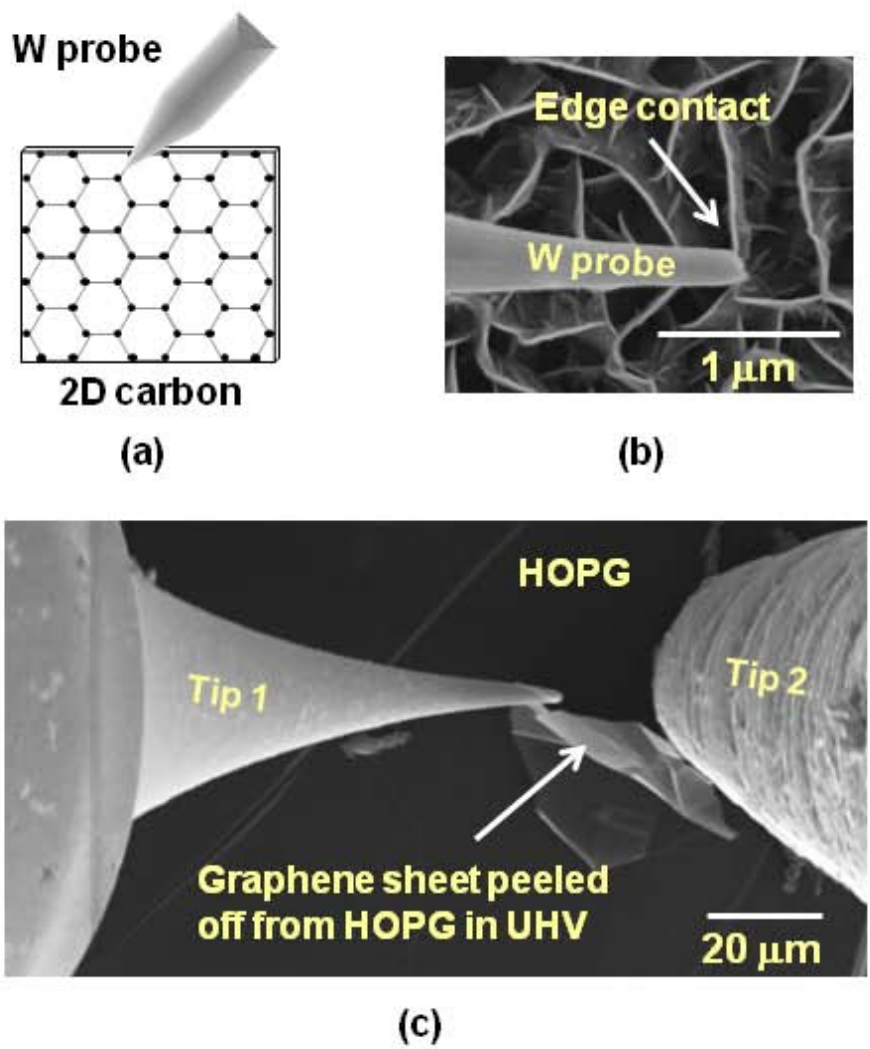

Fig.1

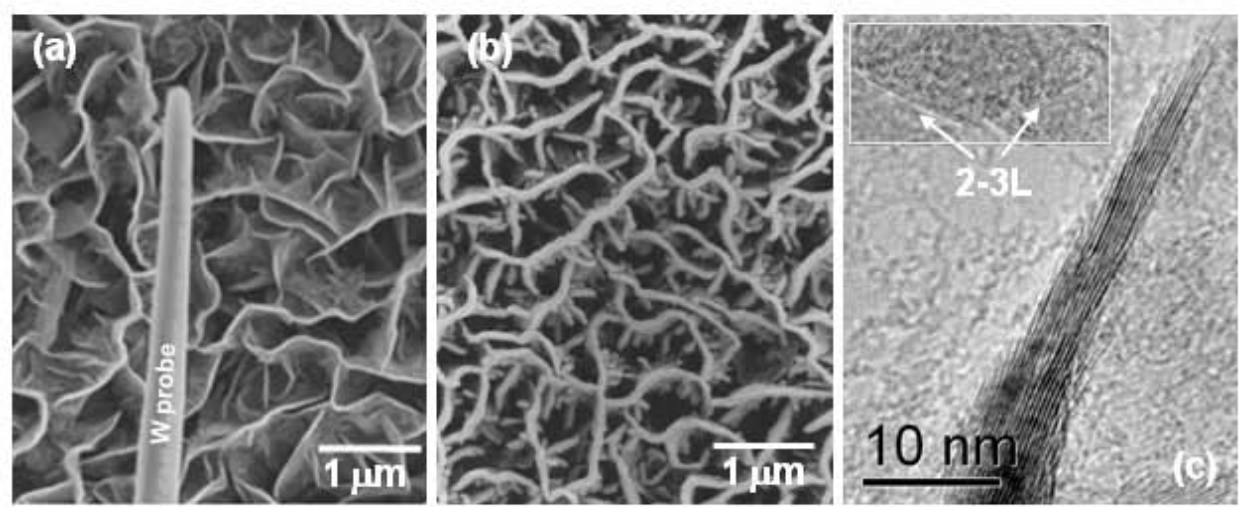

Fig.2

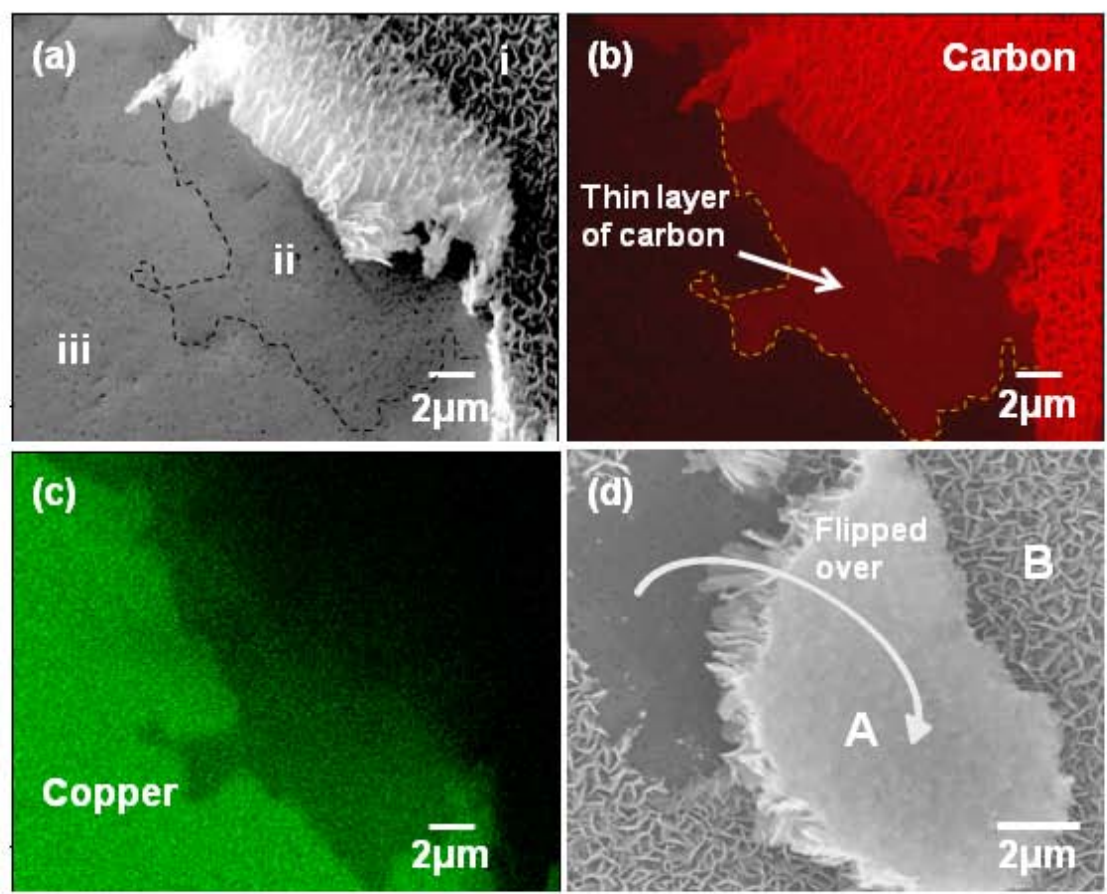

Fig.3

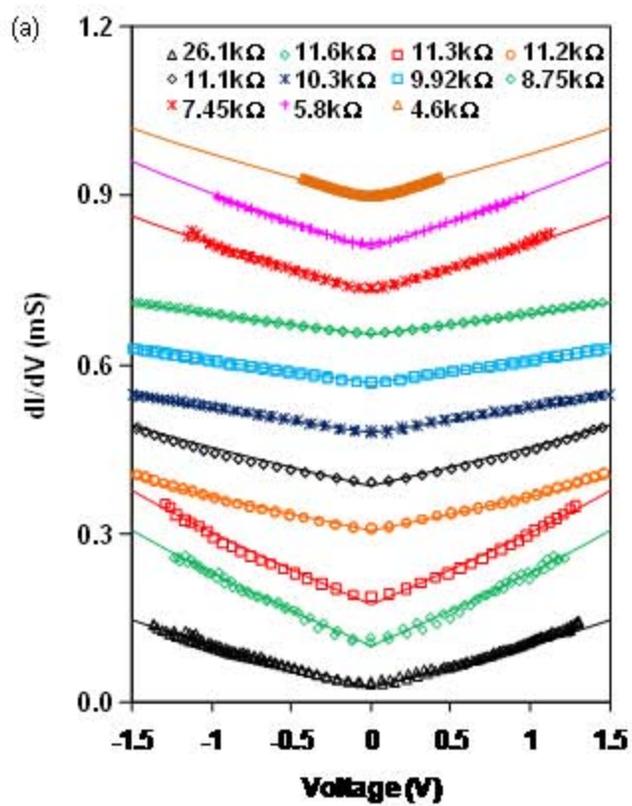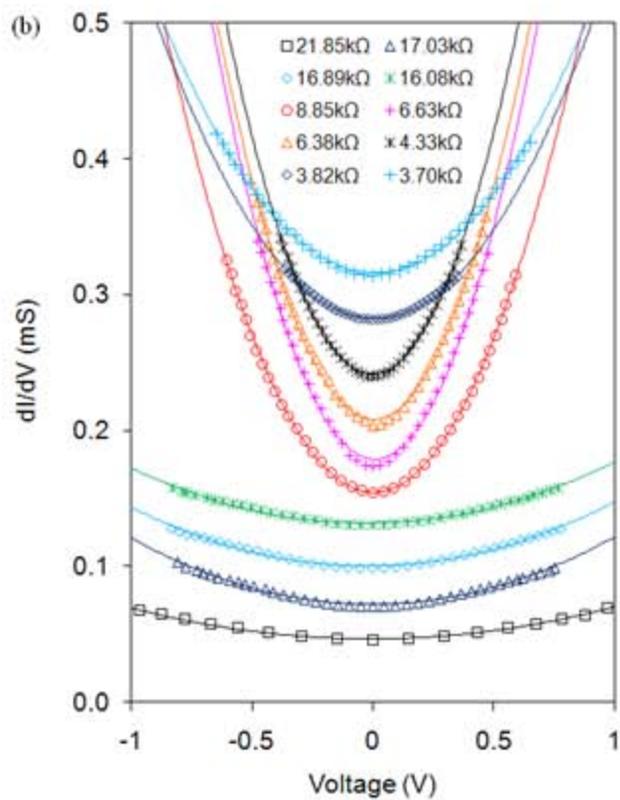

Fig.4

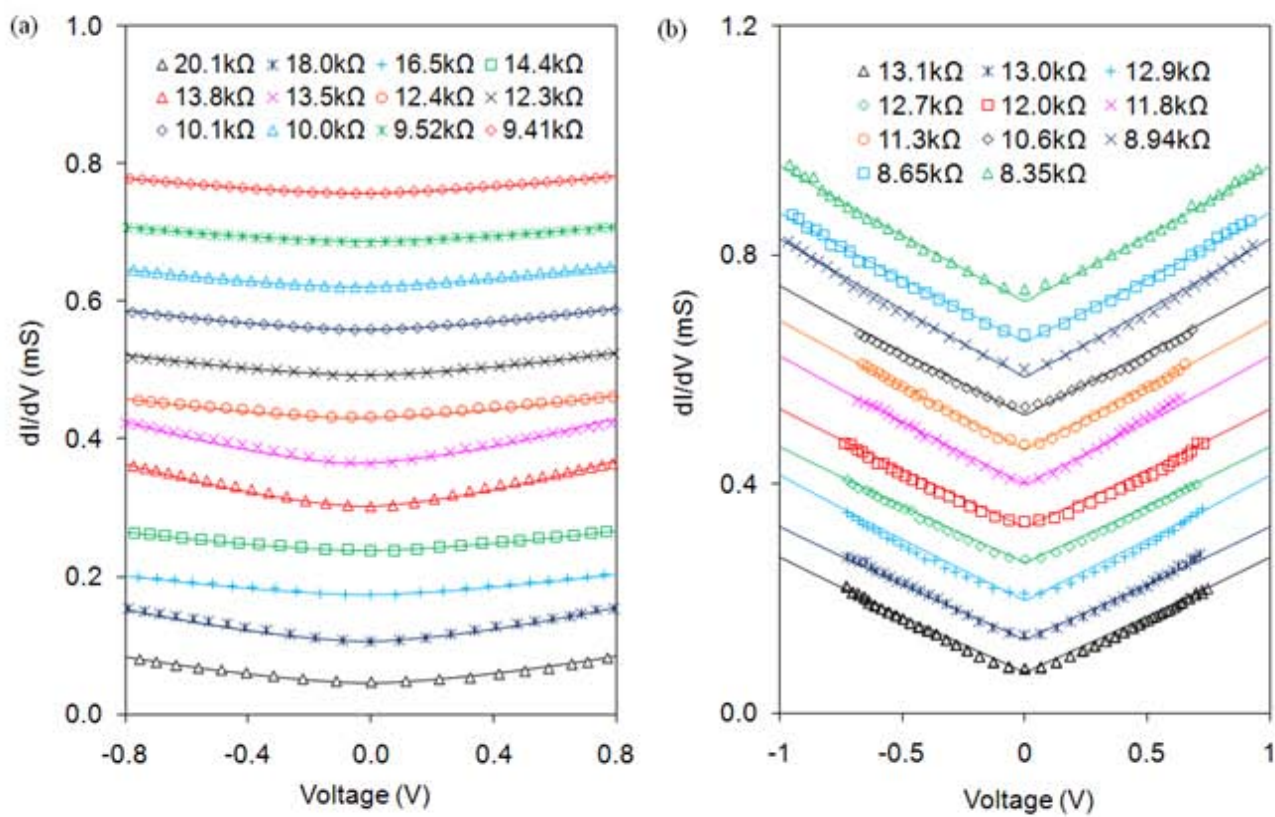

Fig.5

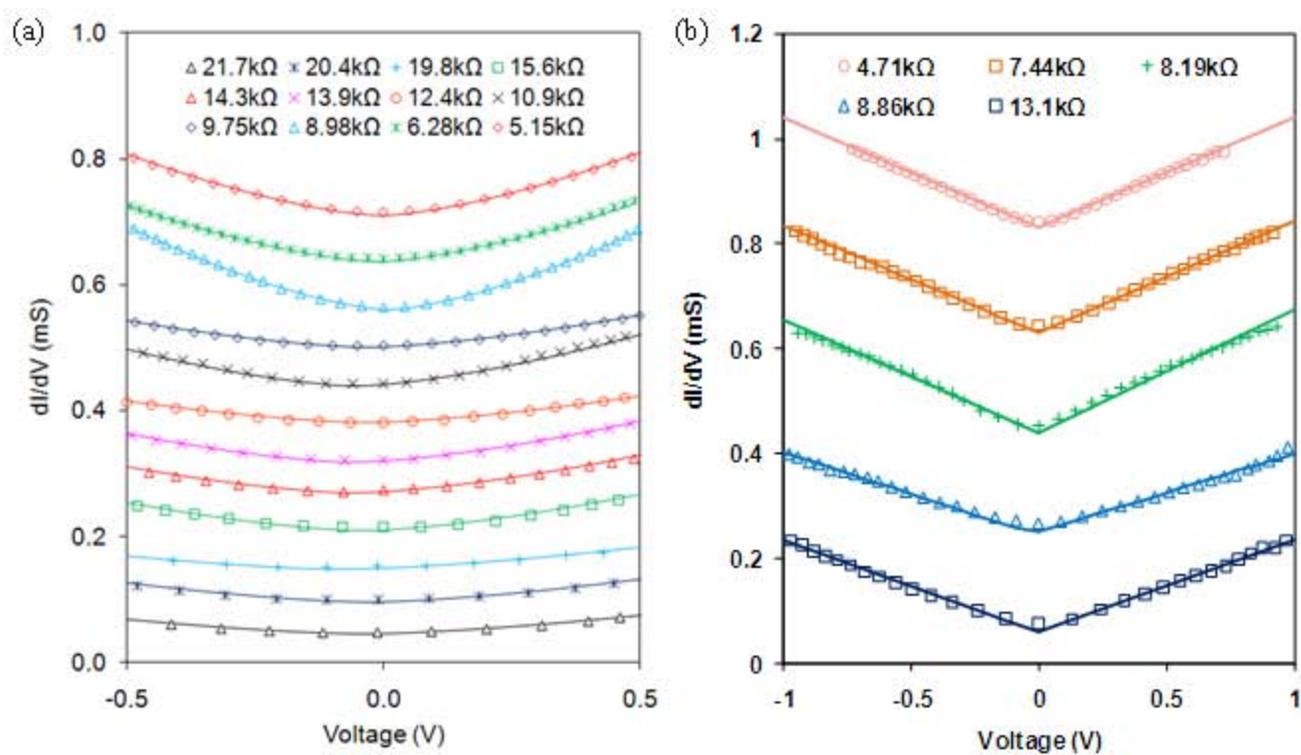

Fig.6

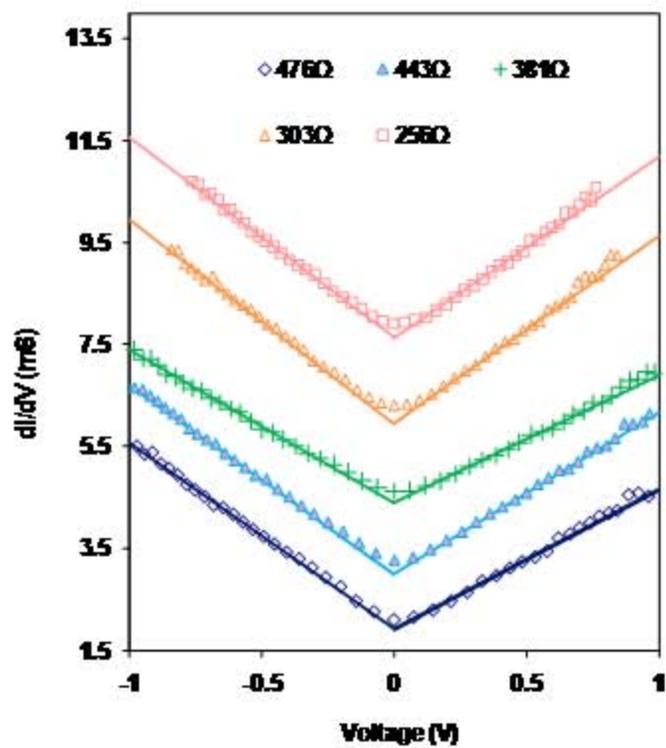

Fig.7

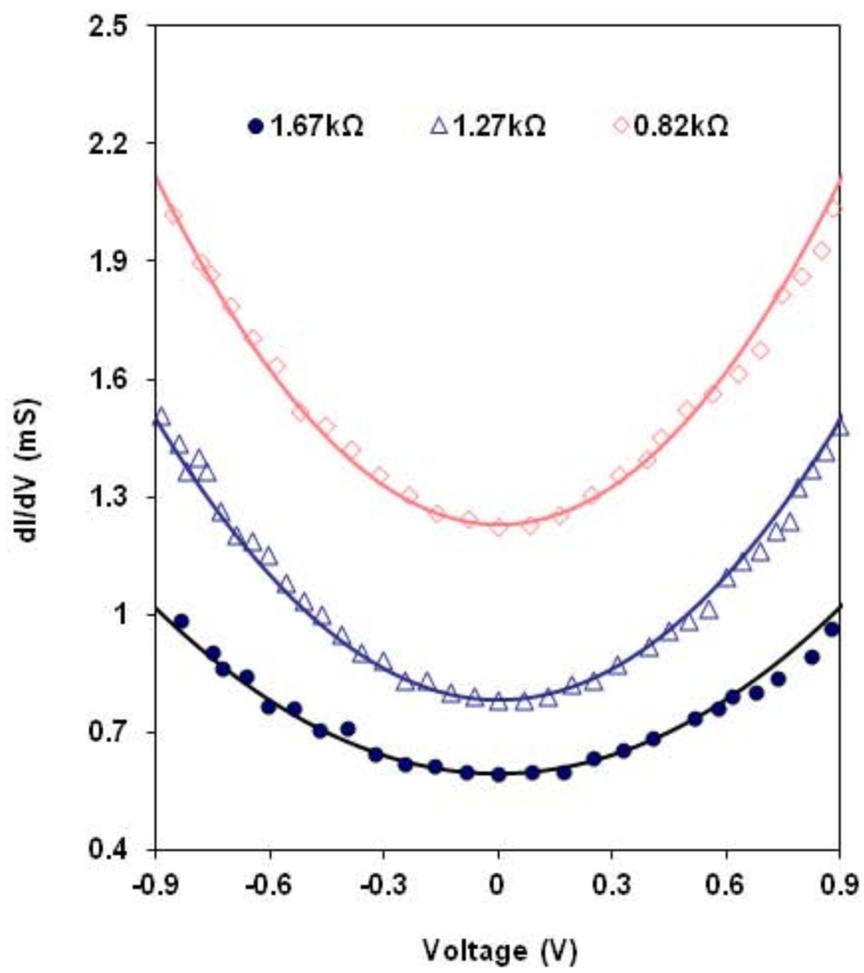

Fig.8

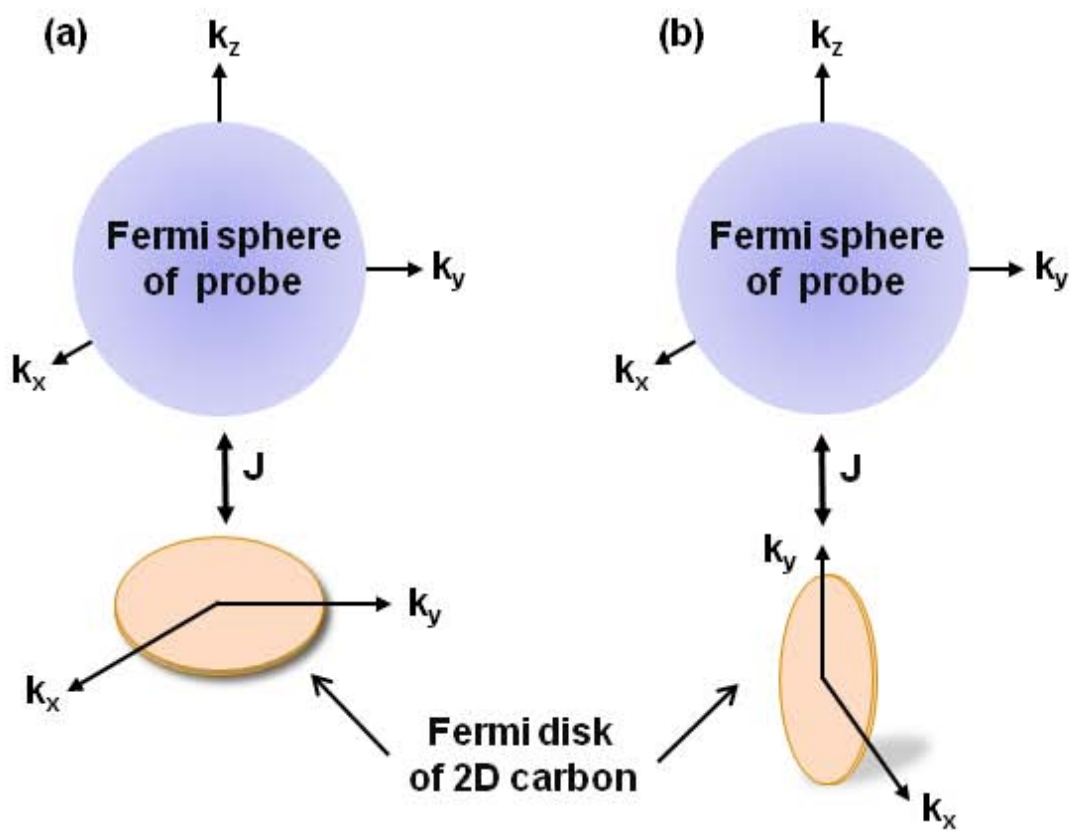

Fig.9

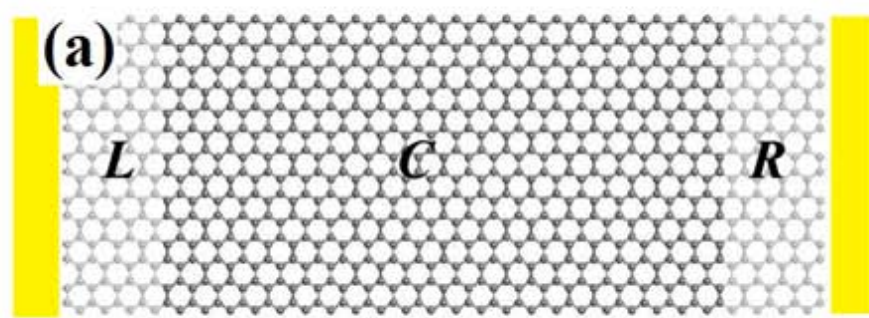
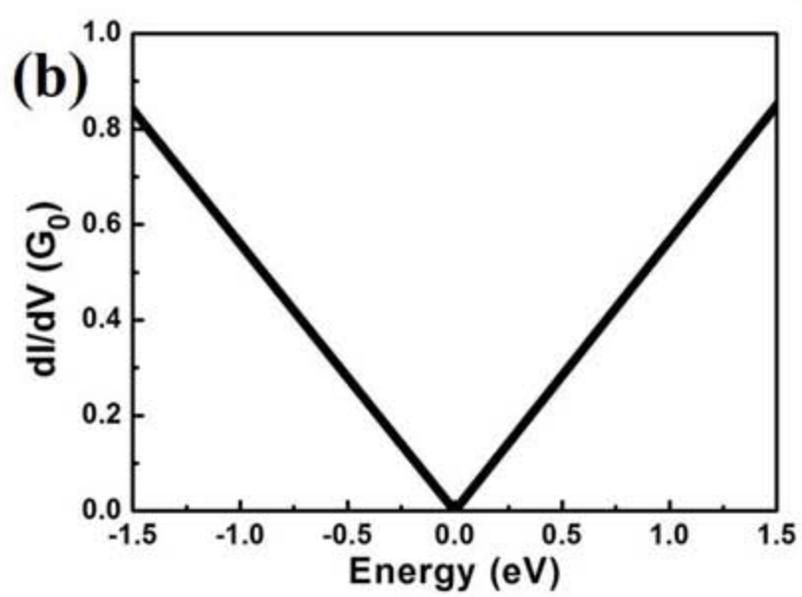

Fig.10

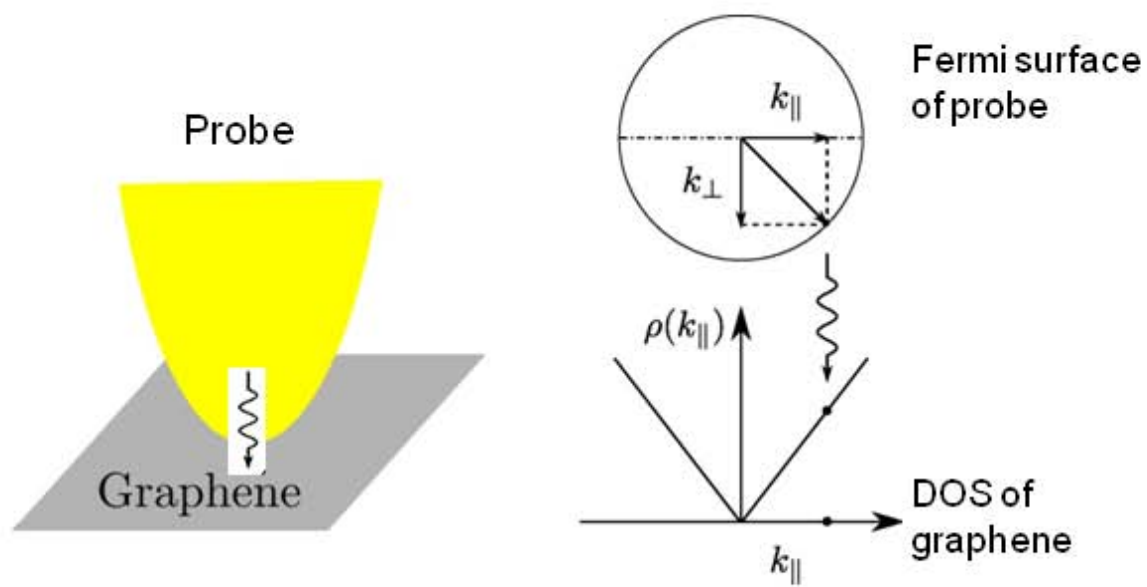

**Fig.11**